\PassOptionsToPackage{table,xcdraw,dvipsnames}{xcolor}
\documentclass[sigconf]{acmart}

\usepackage{amsmath}
\usepackage{booktabs}
\usepackage{arydshln}
\usepackage{multirow}
\usepackage[normalem]{ulem}
\usepackage{pdflscape}
\usepackage{enumitem}

\usepackage{pifont}

\newcommand{\cmark}{\ding{51}}%
\newcommand{\xmark}{\ding{55}}%

\newcommand{\set}[1]{\mathcal{#1}}

\newcommand{\mantis}{\texttt{MANtIS}}
\newcommand{\msdialog}{\texttt{MSDialog}}
\newcommand{\ubuntu}{\texttt{UDC\textsubscript{DSTC8}}}

\newcommand{\bm}{BM25}
\newcommand{\rmthree}{RM3}
\newcommand{\resptocontext}{$resp2ctxt$}
\newcommand{\resptocontextlu}{$resp2ctxt_{lu}$}

\newcommand{\usep}{[U]}
\newcommand{\tsep}{[T]}

\AtBeginDocument{%
  \providecommand\BibTeX{{%
    \normalfont B\kern-0.5em{\scshape i\kern-0.25em b}\kern-0.8em\TeX}}}

\setcopyright{none}

\begin{document}

\title{Sparse and Dense Approaches for the Full-rank Retrieval of Responses for Dialogues}



\author{Gustavo Penha}
\affiliation{%
\institution{Delft University of Technology}
\city{Delft}
\country{Netherlands}}
\email{g.penha-1@tudelft.nl}

\author{Claudia Hauff}
\affiliation{%
\institution{Delft University of Technology}
\city{Delft}
\country{Netherlands}}
\email{c.hauff@tudelft.nl}

\begin{abstract}
  Ranking responses for a given dialogue context is a popular benchmark in which the setup is to re-rank the ground-truth response over a limited set of $n$ responses, where $n$ is typically 10. The predominance of this setup in conversation response ranking has lead to a great deal of attention to building neural \emph{re}-rankers, while the first-stage retrieval step has been overlooked. Since the correct answer is always available in the candidate list of $n$ responses, this artificial evaluation setup assumes that there is a first-stage retrieval step which is always able to rank the correct response in its top-$n$ list. In this paper we focus on the more realistic task of \emph{full-rank retrieval of responses}, where $n$ can be up to millions of responses. We investigate both dialogue context and response expansion techniques for sparse retrieval, as well as zero-shot and fine-tuned dense retrieval approaches. Our findings---based on three different information-seeking dialogue datasets---reveal that a learned response expansion technique is a solid baseline for sparse retrieval. We find the best performing method overall to be dense retrieval with intermediate training---a step after the language model pre-training where sentence representations are learned---followed by fine-tuning on the target conversational data. We also investigate the intriguing phenomena that harder negatives sampling techniques lead to worse results for the fine-tuned dense retrieval models. The code and datasets are available at \textcolor{RubineRed}{\url{https://github.com/Guzpenha/transformer_rankers/tree/full_rank_retrieval_dialogues}}.

  \end{abstract}

\maketitle

\section{Introduction}


\emph{Conversational search} is concerned with creating agents that satisfy an information need by means of a \emph{mixed-initiative} conversation through natural language interaction, rather than through the traditional search engine results page. It is an active area of research---as evident for instance in the SCAI\footnote{\url{https://scai.info/}} workshop series and two recent and independent surveys about the topic~\cite{DBLP:journals/corr/abs-2201-08808,DBLP:journals/corr/abs-2201-05176}---due to the widespread deployment of voice-based agents, such as Google Assistant and Microsoft Cortana. A popular approach to conversational search is retrieval-based~\cite{anand2020conversational}: given an ongoing conversation and a large corpus of historic conversations, retrieve the response that is best suited from the corpus (i.e., conversation response ranking~\cite{wu2017sequential, yang2018response,penha2019curriculum,yang2020iart,han-etal-2021-fine}). This retrieval-based approach does not require task-specific semantics by domain experts~\cite{henderson2019convert}, and it avoids the difficult task of dialogue generation, which often suffers from uninformative, generic responses~\cite{li2016diversity} or responses that are incoherent given the dialogue context~\cite{li2016persona}.

Early neural models for response ranking were based on the interaction of the concatenated dialogue context and the response in a single-turn manner~\cite{lowe2015ubuntu,kadlec2015improved}, with CNN and LSTM architectures. Researchers later explored using multiple interactions, where each utterance in the dialogue context is matched with the response, known as multi-turn matching neural networks~\cite{yan2016learning,wu2016sequential,zhou2018multi,gu2019interactive,lin2020world}. More recently, due to the effectiveness of heavily pre-trained transformer-based language models such as BERT~\cite{devlin2018bert} across NLP tasks, they have become the predominant approach for response ranking~\cite{penha2019curriculum,xu2020learning,whang2021response,zhang2021structural,su2020dialogue,gu2020speaker,whang2019effective}. This is currently the most successful family of methods for retrieval-based chatbots~\cite{zhang2021advances}\footnote{\url{https://github.com/JasonForJoy/Leaderboards-for-Multi-Turn-Response-Selection}}.

The offline evaluation of such neural ranking models is to rank the ground-truth response over a limited set of $n$ responses and measure the number of relevant responses found in the first $K$ positions--- $Recall_n@K$~\cite{zhang2021advances}. Since the entire collection of available responses is typically way bigger\footnote{While for most benchmarks~\cite{zhang2021advances} we have only 10--100 candidate responses, a working system with the Reddit dataset from PolyAI \url{https://github.com/PolyAI-LDN/conversational-datasets} for example would need to retrieve from 3.7 billion responses.} than such set of candidates, this setup is in fact a re-ranking problem, where we have to select the best response out of a few options. Additionally, in existing benchmarks the correct response is traditionally amongst the $n$ responses to re-rank~\cite{penha2020challenges}. This is thus an artificial evaluation that overlooks the first-stage retrieval step, which needs to retrieve the $n$ responses that will be later re-ranked. If the first-stage model, e.g. BM25, fails to retrieve relevant responses, the \emph{retrieve then re-rank} pipeline will also fail even if the re-ranker is an oracle model.




In this paper we offer a novel and comprehensive comparison of supervised and unsupervised, dense and sparse retrieval models\footnote{Although we evaluate them as standalone methods for the full-rank retrieval problem, they can also be employed as first-stage retrievers followed by a re-ranking step.} for the overlooked problem of \emph{full-rank retrieval of responses for dialogues}. We adapt prominent techniques for the problem, i.e. effective in other ranking tasks such as passage retrieval, including document expansion for the task of ranking responses for dialogue contexts. 

We provide here empirical evidence to the following open questions when setting up a full-rank retrieval system for conversation response ranking. What is the effectiveness of sparse and dense retrieval when ranking responses from the \emph{entire collection}? How do dense models compare with strong sparse baselines? What is their effectiveness in a zero-shot setup? What is the effect of adding an intermediate representation learning step between the language model pre-training and the training with conversational data?

We also shed light on the important problem of selecting negative samples when training dense retrieval models, which has been shown to have a great effect on the final effectiveness in different retrieval tasks~\cite{xiong2020approximate,zhan2021optimizing}. Unlike previous work that study sampling out of a few random conversational responses in the re-ranking setup through modifications to a cross-encoder model~\cite{li2019sampling}, we study the harder problem of sampling negative responses from the entire collection. We are the first to investigate different hypotheses in the context of negative sampling of responses for dialogues that can explain difficulties in using harder negatives in the training of dense retrievers.





Our main findings in building retrieval models of responses for dialogues in the full-rank setting are:

\begin{itemize}
    \item While dialogue context expansion is not successful for sparse retrieval, supervised response expansion through the proposed \resptocontextlu{} is a strong baseline for full-rank retrieval of responses for dialogues.
    \item Dense retrieval without access to the target dialogue data, i.e. the zero-shot scenario, is able to beat a strong sparse baseline only when it has access to a large amount of out of domain supervision data.
    \item Dense retrieval models that have intermediate training followed by fine-tuning with the target data are the best performing models, even with a simple random sampling approach for obtaining negative responses.
    \item Harder negative sampling techniques lead to worse effectiveness. We found evidence for the hypothesis that false positives strongly contribute to this phenomena. Denoising is an effective approach to take advantage of harder negative samples.
\end{itemize}

\section{Related Work}
In this section we analyze previous work pertinent to this paper by first discussing current research in the domain of retrieval-based chatbots, followed by reviewing (un)supervised dense and sparse retrieval approaches. We conclude this section with an overview of findings on the topic of selecting negative samples to train neural ranking models.

\subsection{Ranking Responses for Dialogues}

Early neural models for response ranking were based on matching the representations of the concatenated dialogue context and the representation of a response in a single-turn manner with architectures such as CNN and LSTM~\cite{lowe2015ubuntu,kadlec2015improved}. Researchers later explored matching each utterance in the dialogue context with the response with more complex neural architectures~\cite{yan2016learning,wu2016sequential,zhou2018multi,gu2019interactive,lin2020world}.

Using heavily pre-trained language models for ranking was first shown to be effective by \citep{nogueira2019passage}. They used a BERT model to re-rank the responses of a first stage retrieval system on the \texttt{MSMarco} passage retrieval task and showed significant improvements in effectiveness. Such language models for ranking have quickly became a predominant approach in information retrieval~\cite{lin2021pretrained}. This was also shown to be effective for ranking responses in conversations~\cite{whang2019effective,penha2019curriculum}.~\citet{penha2019curriculum} showed a way of using a BERT-based re-ranking model for the dialogues domain, which is improved when notions of difficulty are taken into account in a curriculum learning training procedure. 

One limitation of transformer-based language models is that they do not take into account the structure of a dialogue. ~\citet{gu2020speaker} proposed adding another embedding layer to BERT that takes into account the speaker of the dialogue. Dialogue-aware training has also been further explored, for example both by~\citet{han-etal-2021-fine} and~\citet{whang2021response} who proposed different modifications to the conversational data to improve the fine-tuning of language models. Building better re-ranking models for dialogue tasks is still a very active research field as seen by recent surveys on the topic~\cite{tao2021survey,zhang2021advances}.

In contrast, full-rank retrieval of responses has been under explored~\cite{penha2020challenges}. ~\citet{lan2021exploring} showed that a BERT-based dense retrieval model outperforms BM25 on the full-rank task.~\citet{tao2021building} later proposed a mutual learning model that trains both the dense retrieval bi-encoder model and the cross-encoder re-ranker model at the same time. They also showed that such dense model is more effective than BM25 without expansion techniques for the full-rank problem of retrieving responses for dialogues. 

A limitation of previous work is that a strong sparse retrieval baseline model, e.g. BM25+\textit{dialogue context expansion} or BM25+\textit{response expansion}, was not compared. Such expansion methods are capable of combating the problem of vocabulary mismatch and thus the question if dense models are able to outperform sparse ones when using expansion techniques is still unanswered. We expand on the analysis of previous work~\cite{lan2021exploring,tao2021building} by looking into stronger sparse baselines, evaluating the effect of intermediate training, testing zero-shot effectiveness of dense models and studying the effect of other negative sampling methods besides random sampling.



\subsection{Dense and Sparse Retrieval}

The proposed conceptual framework by \citep{lin2021proposed} argues for a categorization of retrieval models into two dimensions: supervised vs. unsupervised and dense vs. sparse representations\footnote{A distinction can also be made of cross-encoders and bi-encoders, where the first encode the query and document jointly as opposed to separately~\cite{thakur2020augmented}. Cross-encoders are typically applied in a re-ranking step due to their inefficiency and thus are not the focus of this paper.}. An \emph{unsupervised} sparse representation model such as BM25~\cite{robertson1994some} and TF-IDF~\cite{jones1972statistical} represents each document and query with a sparse vector with the dimension of the collection's vocabulary, having many zero weights due to non-occurring terms. Since the weights of each term are calculated using term statistics they are considered unsupervised methods. 


A \emph{supervised} sparse retrieval model such as COIL~\cite{gao2021coil}, SPLADE~\cite{formal2021splade}, TILDE~\cite{zhuang2021fast} and DeepImpact~\cite{mallia2021learning} can take advantage of the effectiveness of transformer-based language models by changing the terms' weights from collection statistics to something that is learned. DeepCT~\cite{dai2019context} for example learns term weights with a transformer-based regression model from the supervision of the \texttt{MSMarco} dataset. Approaches that only modify non-zero weights however are not able to address the vocabulary mismatch problem~\cite{furnas1987vocabulary}, as non-zero terms will not be affected. One way to address such problem in sparse retrieval is by using query expansion methods. RM3~\cite{abdul2004umass} has been shown to be a competitive query expansion technique that uses pseudo-relevance feedback to add new terms to the queries followed by another final retrieval step using the modified query.

Document expansion has also been shown to be an effective technique to improve sparse retrieval, which is able to address the vocabulary mismatch problem. The core idea is to create pseudo documents that have expanded terms and use them instead when doing retrieval. Doc2query~\cite{nogueira2019doc2query} is an effective approach to document expansion that uses a language model to predict the queries which might be issued to find the document. The predictions of this model are used to create the augmented pseudo documents. Expansion techniques are able to modify non-zero weights by adding terms that did not exist in the query or document. 

Supervised dense retrieval models, such as ANCE~\cite{xiong2020approximate}, RocketQA~\cite{qu2020rocketqa}, PAIR~\cite{ren2021pair} and coCodenser~\cite{gao2021unsupervised}, represent query and documents in a smaller fixed-length space, for example of 768 dimensions, which can naturally capture semantics. In this manner they are able to address the vocabulary mismatch problem. While dense retrieval models have shown to consistently outperform BM25 in multiple datasets, this is not so easily the case when dense retrieval models do not have access to training data from the target task, known as the \emph{zero-shot scenario}. The \texttt{BEIR} benchmark~\cite{thakur2021BEIR} showed that BM25 was superior to dense retrieval from 9--18 (depending on the model) out of the 18 datasets under this evaluation scheme. While the zero-shot scenario offers a fairer comparison of dense models with unsupervised sparse models, learned dense retrieval models should also be compared with learned sparse models, e.g. BM25+doc2query.

Unlike previous work that compares supervised and unsupervised, dense and sparse retrieval models for other tasks such as passage ranking, we provide a novel and comprehensive comparison for the problem of \emph{full-rank retrieval of responses for dialogues}.


\subsection{Negative Sampling for Ranking}

Neural ranking models are known to require a reasonable amount of labeled queries and documents for training. The \texttt{TREC-DL} dataset~\cite{craswell2020overview}---created from \texttt{MSMarco}~\cite{nguyen2016ms}---offers a large training set, 367k queries, with \emph{no negative labels} and often only one positive label per query. This is also the case for conversation response ranking datasets, such as \mantis{}~\cite{penha2019introducing} and \ubuntu{}~\cite{Kummerfeld_2019}, where we have a large number of training data consisting of relevant responses for dialogue contexts and no explicitly negative labelled responses.

In order to train neural ranking models, negative candidates are also necessary since it is prohibitively expensive to use every other document in the collection as negative sample for a query, and manual relevance judgements are expensive to obtain for a large scale dataset. This motivates automatically finding non-relevant documents for a combination of query and relevant documents, known as \emph{negative sampling}.

This problem is also present for other machine learning techniques that do self-supervision through contrastive learning in different domains such as computer vision, natural language and graphs~\cite{robinson2020contrastive,yang2020understanding,jiang2021lightxml}. For example the well-known \textit{word2vec}~\cite{mikolov2013efficient} word embedding technique randomly samples words that are not relevant for the context (other words in the sentence) to distinguish from the actual word that is part of the context.

In information retrieval, since most of the documents in a collection are not relevant for a given query, a simple approach is to obtain negative candidates through randomly selecting documents (the same way \textit{word2vec} selects random words), with the exception of the documents that were labeled relevant. A popular technique is to use in-batch negative samples, which are in essence random and make the training procedure efficient~\cite{humeau2019poly,mazare2018training,lan2021exploring}. However, a limitation of random samples is that the documents might be too easy for the ranking model to discriminate from relevant ones, while for negative documents that are harder to distinguish the model might still struggle.

For this reason, another popular approach has been to use a ranking model to retrieve negative documents using the given query with a classical retrieval technique such as BM25. This leads to finding negative documents that are closer to the query in the sparse representation space, and thus \emph{harder} negatives. 



Since dense retrieval models have been outperforming sparse retrieval in a number of cases with available training data, more complex negative sampling techniques taking advantage of dense retrieval models' effectiveness have been proposed. The ANCE~\cite{xiong2020approximate} model uses the dense model itself to find negatives (ANCE negatives), which is asynchronously updated in checkpoints. This effectively makes the model find harder and harder negatives throughout training.~\citet{gao2021rethink} proposed to localize the training procedure with the same negative samples from the target distribution of the target set used for evaluation.~\citet{hofstatter2021efficiently} trained dense models using balanced batches, where an equal number of easy and difficult negative---according to the distillation margins---documents can be found.

Negative sampling has also been given attention in the training of cross-encoder re-ranking for conversation response ranking.~\citet{li2019sampling} showed that if different selection methods to obtain negatives from a pool of randomly sampled responses are employed, re-rankers can be trained more effectively. Note that this is quite different to negative sampling from the entire collection to train bi-encoders which is what we are focused on here. ~\citet{qiu2021challenging} showed that with negatives generated with a pre-trained language model such as DialogGPT~\cite{zhang2019dialogpt}, which can also be seen as a data augmentation technique, one can obtain effectiveness improvements for re-ranking responses to dialogues. The combination of augmented negatives through different negative sampling techniques has also been shown to be effective for re-ranking responses when using a multi-level ranking loss~\cite{lin2020world}. 

Building on these prior works, we are the first to investigate different hypotheses in the context of negative sampling of responses for dialogues that might explain difficulties in using harder negatives in the training of dense retrievers for the full-rank task.


\section{Full-rank Retrieval for Dialogues}

\begin{figure*}[ht!]
    \centering
    \includegraphics[width=.99\textwidth]{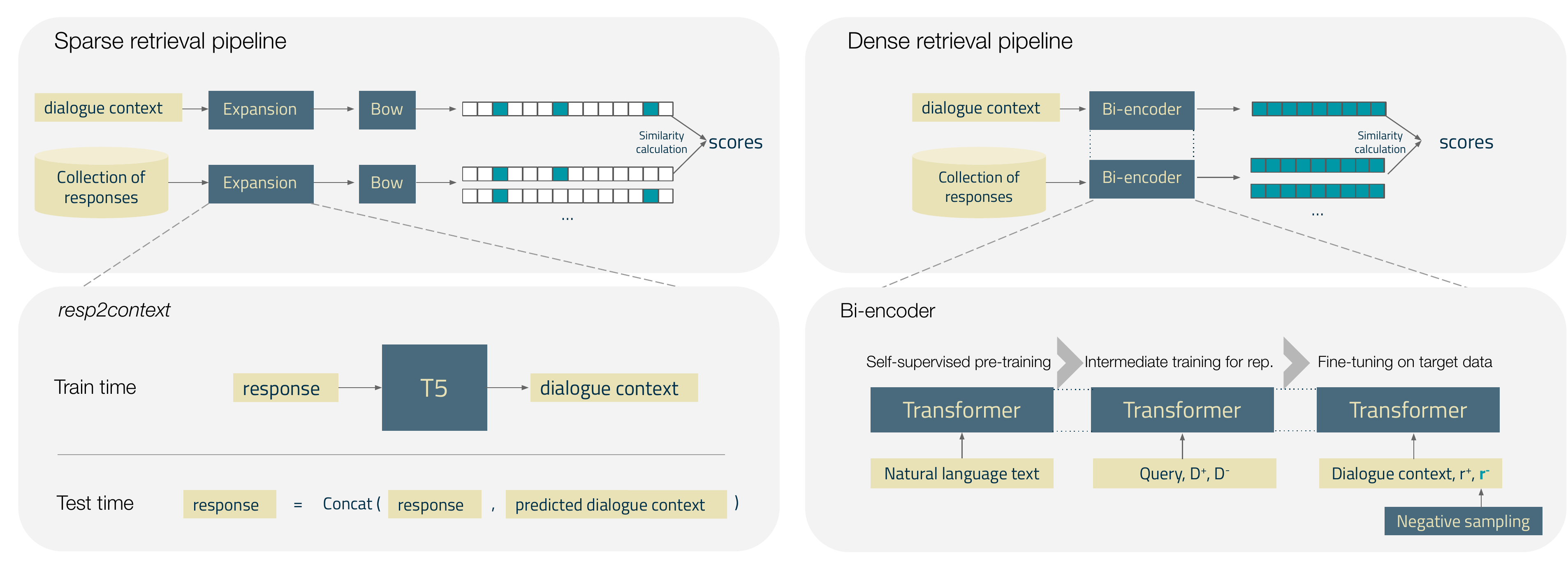}
    \caption{Pipelines for sparse and dense retrieval for the task of full-rank retrieval of responses for dialogues.}
    \label{fig:methods_pipelines}
\end{figure*}

In this section we first describe the problem of full-rank retrieval of responses, followed by the proposed sparse and then dense approaches for the problem. A summary of both pipelines can be seen in Figure~\ref{fig:methods_pipelines}.

\subsection{Problem Definition}
The task of full-rank retrieval of responses for dialogues, concerns retrieving the best response out of the entire collection given the dialogue context. Formally, let $\set{D}=\{(\set{U}_i, \set{R}_i, \set{Y}_i)\}_{i=1}^{M}$ be a data set consisting of $M$ triplets: dialogue context, response candidates and response relevance labels. The dialogue context $\set{U}_i$ is composed of the previous utterances $\{u^1, u^2, ... , u^{\tau}\}$ at the turn $\tau$ of the dialogue. The candidate responses $\set{R}_i = \{r^1, r^2, ..., r^n\}$ are either ground-truth responses $r^+$ or negative sampled candidates $r^-$, indicated by the relevance labels $\set{Y}_i = \{y^1, y^2, ..., y^n\}$.

In previous work, the number of candidates is limited, typically $n = 10$. Since we are concerned with the full-rank task and not the re-ranking setting, in all of our experiments we have $n$ as the size of the responses available in the collection. By design the number of ground-truth responses is usually one, the observed response in the conversational data\footnote{One limitation of current benchmarks is that it is likely to exist many semantically equivalent responses as well as different relevant responses to the same dialogue context, while the ground-truth ones are considered to be single responses given by other users.}.

The task is then to learn a ranking function $f(.)$ that is able to generate a ranked list for the set of candidate responses $\set{R}_i$ based on their predicted relevance scores $f(\set{U},r)$.

To evaluate the effectiveness of the retrieval systems we use $R@K$, and thus we evaluate the models effectiveness in finding the correct response out of the whole possible set of responses at position $K$.

\subsection{Sparse Retrieval}

In order to do sparse retrieval of responses we rely on classical retrieval methods with query and document expansion techniques. One of the limitations of sparse retrieval is that, since it represents each dialogue context and response using the existing terms in a bag-of-words manner, the vocabulary mismatch problem might occur. Such expansion techniques are able to overcome this problem if they append new words to the dialogue contexts and responses.

For this reason we propose here to do \emph{dialogue context expansion} with RM3~\cite{abdul2004umass}, which is a very competitive unsupervised method that assumes that the top-ranked responses by the sparse retrieval model are relevant. From such pseudo-relevant responses words are selected and an expanded dialogue context is generated, which will then be used by the sparse retrieval method to generate the final ranked list.

In order to expand the responses to be retrieved, we propose here \resptocontext{}. This is an adaptation of the popular and effective \textit{doc2query}~\cite{nogueira2019doc2query} approach for dialogues. Formally, we fine-tune a generative transformer model for the task of generating the dialogue context $\set{U}_i$ from the ground-truth response $r^+$. This model is then used to generate expansions for all responses in the collection. They are appended to the responses and the sparse retrieval method itself is not modified.

This way, \resptocontext{} allows for two things: term re-weighting (adding terms that already exist in the document) and addition of new terms (to deal with the vocabulary mismatch problem).

Unlike most adhoc retrieval problems where the queries are smaller than the documents, full-rank retrieval of responses for dialogues is the exact opposite. For example, while the TREC-DL 2020 passage and document retrieval tasks the queries have between 5--6 terms on average and the passages and documents have over 50 and 1000 terms respectively, the dialogue contexts (queries) have between 70 and 474 terms on average depending on the dataset while the responses (documents) have between 11 and 71 terms on average, as seen in the first two rows of Table~\ref{table:resp2context_stats}. This is a challenge for the generative model, since generating larger pieces of text is a more difficult problem than smaller ones, e.g. more room for error.

Motivated by this, we also explored an adaptation of \resptocontext{} that aims to generate only the last utterance of the dialogue context: \resptocontextlu{}. This model is trained to generate $u^{\tau}$ from $r+$. The underlying premise is that the part that needs to be answered by the dialogue context is the last utterance, and if this is correctly generated by \resptocontextlu{}, the sparse retrieval method will be able to find the correct response from the collection.

\subsection{Dense Retrieval}

In order to do dense retrieval of responses we rely on methods that learn to represent the dialogue context and the responses separately in a dense embedding space. This is later used to rank responses by their similarity to the dialogue context. We rely here on heavily pre-trained language transformer models, such as BERT~\cite{devlin2018bert}, RoBERTa~\cite{liu2019roberta} or MPNet~\cite{song2020mpnet}, to obtain such representations of the dialogue context and response. This approach is generally referred as a \emph{bi-encoder} model~\cite{lin2021pretrained}.

\subsection{Intermediate Training}

The first step of the pipeline is to train the representations of the language model with intermediate\footnote{We differentiate this intermediate step to a pre-training step due to the fact that the transformer based language models were already pre-trained for their respective language modelling tasks. For example, BERT is pre-trained for next sentence prediction and masked language modeling and can be later trained to represent queries and documents.} data that does not contain the target domain data. Such intermediate data contains triplets of query, relevant document and negative document and can include multiple datasets. The main advantage of adding this step before fine-tuning the bi-encoder for the target conversational data is to reduce the gap between the pre-training, often including language modelling, and downstream task at hand.

The intermediate training step learns to represent texts (query and documents) by doing a mean pooling function over the transformer final layer, which is then used to calculate the dot-product similarity. The relevant document representation is used to contrast with the representations of the document that is not relevant. Such procedure learns better text representations than a naive approach of simply using the $[CLS]$ token representation of BERT for the dialogue contexts and responses~\cite{reimers-2019-sentence-bert,aghajanyan2021muppet}.

The loss function employs multiple negative texts to learn the representations in a constrastive manner, also known as in-batch negative sampling. This model is then able to do zero-shot retrieval for the full-rank retrieval of responses to dialogue contexts, since it does not have access to the target domain data.

The function $f(\set{U},r)$ can be defined as $dot(\eta(concat(\set{U})),\eta(r))$, where $\eta$ is the representation obtained with the mean pooling of all the output vectors of the transformers language model, and $concat(\set{U}) = u^1 \; | \; \usep \; | \; u^2 \; | \;  \tsep \; | \; ... \; | \; u^{\tau} \;$, where $|$ indicates the concatenation operation. The utterances from the context $\set{U}$ are concatenated with special separator tokens $\usep$ and $\tsep$ indicating end of utterances and turns\footnote{The special tokens $\usep$ and $\tsep$ will not have any meaningful representation in the zero-shot setting, but they can be learned on the fine-tuning step.}.

\subsection{Fine-tuning}

The second step in the pipeline is to fine-tune the model with data from the target domain, in our case dialogue contexts and responses. Since we do not have labeled negative responses, and only relevant ones, the remaining responses can be thought of as non relevant for the dialogue context. Computing the probability of the correct response over all other responses in the dataset would give us $P(r \mid \set{U})=\frac{P(\set{U}, r)}{\sum_{k} P\left(\set{U}, r_{k}\right)}$. Since this computation is prohibitively expensive to calculate, we approximate it using only a few negative samples retrieved by a negative sampling approach. 

The \emph{negative sampling} task is then to: given the dialogue context $\set{U}$ find challenging responses that are not relevant. This can be seen as a retrieval task as well, where one can use a retrieval model to find negatives by applying $f(\set{U}, r)$ for every $r$ in the collection, sorting, and removing $r^{+}$ from the resulting top negatives.

In our experiments we test a number of different approaches to obtain hard negative samples: random sampling, BM25 sampling, bi-encoder sampling and generative transformer models\footnote{The generative approach goes directly from the dialogue context to the negative candidate: $\set{U} \xrightarrow{}  r^{-}$.}.

With such dataset at hand, we continue the training---after the intermediate step---in the same manner as done by the intermediate training step, with the following cross-entropy loss function\footnote{We refer to this loss as MultipleNegativesRankingLoss.} for a batch with size $B$:


\vspace{3mm} 
$\begin{aligned} \mathcal{J}(\mathbf{\set{U}}, \mathbf{r}, \theta) &=-\frac{1}{B} \sum_{i=1}^{B}\left[f\left(\set{U}_{i}, r_{i}\right)-\log \sum_{j=1, j!=i}^{B} e^{f\left(\set{U}_{i}, r_{j}\right)}\right] \end{aligned},$
\vspace{3mm}

where $f(\set{U},r)$ is the dot-product of the mean pooled representation of the transformer model. 


\section{Experimental Setup}
In order to compare the different sparse and dense approaches we consider three large-scale information-seeking conversation datasets\footnote{\msdialog{} is available at~\url{https://ciir.cs.umass.edu/downloads/msdialog/}; \mantis{} is available at~\url{https://guzpenha.github.io/MANtIS/}; \ubuntu{} is available at ~\url{https://github.com/dstc8-track2/NOESIS-II}.} that allow the training of neural ranking models for conversation response ranking: \textbf{\msdialog{}}~\cite{qu2018analyzing} contains 246K context-response pairs, built from 35.5K information seeking conversations from the Microsoft Answer community, a QA forum for several Microsoft products; \textbf{\mantis{}}~\cite{penha2019introducing} contains 1.3 million context-response pairs built from conversations of 14 Stack Exchange sites, such as \textit{askubuntu} and \textit{travel}; \textbf{\ubuntu{}}~\cite{Kummerfeld_2019} contains 184k context-response pairs of disentangled Ubuntu IRC dialogues. 

\subsection{Implementation Details}
For BM25 and BM25+RM3 we rely on the pyserini implementations~\cite{Lin_etal_SIGIR2021_Pyserini}. In order to train \resptocontext{} expansion methods we rely on the Huggingface transformers library~\cite{wolf2019huggingface}, using the \texttt{t5-base} model. We fine-tune the T5 model for 2 epochs, with learning rate of 2e-5, weight decay of 0.01 and batch size of 5. When augmenting the responses with \resptocontext{} we follow docT5query~\cite{nogueira2019doc2query} and append three different context predictions, using sampling and keeping the top-10 highest probability vocabulary tokens.

For the zero-shot dense retrieval models, we rely on the SentenceTransformers~\cite{reimers-2019-sentence-bert} model releases\footnote{The pre-trained models can be found here \url{https://www.sbert.net/docs/pretrained_models.html}}. The library uses Hugginface transformers for the pre-trained models such as BERT~\cite{devlin2018bert}, RoBERTa~\cite{liu2019roberta}, MPNet~\cite{song2020mpnet}. When fine-tuning the dense retrieval models, we rely on the \textit{MultipleNegativesRankingLoss}, which accepts a number of hard negatives, and also uses the remaining in-batch random negatives to train the model. We use a total of 10 negative samples for dialogue context.

We fine-tune the dense retrieval models for a total of 10k steps, and every 100 steps we evaluate the models on a re-ranking task that selects the relevant response out of 10 responses. We use the re-ranking validation MAP to select the best model from the whole training to use in evaluation. We use a batch size of 5, with 10\% of the training steps as warmup-steps. The learning rate is set to 2e-5 and weight decay of 0.01. We use the dot-product of the mean pooled representations for both training and also for test (using FAISS~\cite{johnson2019billion}).

On the follow-up experiments to investigate negative sampling approaches, we denoise negatives on E2 using lists of 100 responses and keeping the bottom 10 as negatives. We expand the collection with an external corpus for E5 using ConvoKit~\cite{chang2020convokit}. We choose datasets which have similar topics to the information-seeking datasets we use\footnote{They are namely \texttt{movie-corpus}, \texttt{wiki-corpus}, \texttt{subreddit-Ubuntu}, \texttt{subreddit-microsoft}, \texttt{subreddit-apple}, \texttt{subreddit-Database}, \texttt{subreddit-DIY}, \texttt{subreddit-electronics}, \texttt{subreddit-ENGLISH}, \texttt{subreddit-gis}, \texttt{subreddit-Physics}, \texttt{subreddit-scifi}, \texttt{subreddit-statistics}, \texttt{subreddit-travel} and \texttt{subreddit-worldbuilding}}, amounting to a total of 17M non-empty candidate responses. For experiment E6 we generate the negative candidates using Hugginface~\cite{wolf2019huggingface} conversational pipelines, with the pre-trained models \texttt{DialoGPT-large} and \texttt{blenderbot-400M-distill}.

\subsection{Evaluation}
To evaluate the effectiveness of the retrieval systems, instead of resorting to the standard evaluation metric in conversation response ranking~\cite{yuan2019multi,gu2020speaker,tao2019multi} which is recall at position $K$ with $n$ candidates\footnote{For example $R_{10}@1$ indicates the number of relevant responses found at the first position when the model has to rank 10 candidate responses.} $R_n@K$, we set $n$ to be the entire collection of answers, and thus we evaluate the models effectiveness in finding the correct response out of the whole possible set of responses: $R@K$. We perform Students t-tests at confidence level of 0.95 with Bonferroni correction to compare statistical significance of methods across different dialogue contexts.    
\begin{table*}[ht!]

\caption{Effectiveness of sparse and dense retrieval methods for the full-rank retrieval of responses for dialogues. Bold values indicate the highest recall values for each group (type of approach). Superscripts indicate statistically significant improvements using Students t-test with Bonferroni correction. \textit{$\dagger$=other methods from the same group; $1$=best from unsupervised sparse retrieval ; $2$=best from supervised sparse retrieval; $3$=best from zero-shot dense retrieval. We only do comparisons between the best approaches of each category, and within categories.}}
\label{table:main_table_results}
\begin{tabular}{@{}llrllllll@{}}
\toprule
 &  &  & \multicolumn{2}{c}{\textbf{\mantis{}}} & \multicolumn{2}{c}{\textbf{\msdialog{}}} & \multicolumn{2}{c}{\textbf{\ubuntu{}}} \\ \midrule
 &  &  & \textbf{R@1} & \textbf{R@10} & \textbf{R@1} & \textbf{R@10} & \textbf{R@1} & \textbf{R@10} \\ \midrule
(0) & \multicolumn{2}{l}{Random} & 0.000 & 0.000 & 0.000 & 0.001 & 0.000 & 0.001 \\ \midrule
 & \multicolumn{2}{l}{\textbf{Unsupervised sparse retrieval}} &  &  &  &  &  &  \\ \midrule
(1a) & \multicolumn{2}{l}{\bm{}} & \textbf{0.133$^{\dagger}$} & \textbf{0.299$^{\dagger}$} & \textbf{0.064$^{\dagger}$} & \textbf{0.177$^{\dagger}$} & \textbf{0.027$^{\dagger}$} & \textbf{0.070$^{\dagger}$} \\
(1b) & \multicolumn{2}{l}{\bm{} + \rmthree{}} & 0.073 & 0.206 & 0.035 & 0.127 & 0.011 & 0.049 \\ \midrule
\textbf{} & \multicolumn{2}{l}{\textbf{Supervised sparse retrieval}} &  &  &  &  &  & \\ \midrule
(2a) & \multicolumn{2}{l}{\bm{} + \resptocontext{}} & 0.135 & 0.309 & 0.074 & \textbf{0.208} & 0.028 & 0.067 \\
(2b) & \multicolumn{2}{l}{\bm{} + \resptocontextlu{}} & \textbf{0.147$^{\dagger 1}$} & \textbf{0.325$^{\dagger 1}$} & \textbf{0.075$^{1}$} & 0.202$^{1}$ & \textbf{0.029$^{}$} & \textbf{0.076$^{}$}\\ \midrule
 & \multicolumn{2}{l}{\textbf{Zero-shot dense retrieval}} &  &  &  &  &  & \\
 & \textbf{Model$_{LanguageModel}$} & \textbf{Intermediate data} &  &  &  &  &  \\ \midrule
(3a) & ANCE$_{roberta-base}$ & 600K \texttt{MSMarco-PR} & 0.048 & 0.111 & 0.050 & 0.124 & 0.010 & 0.028 \\
(3b) & TAS-B$_{distillbert-base}$ & 400K \texttt{MSMarco-PR} & 0.062 & 0.143 & 0.060 & 0.157 & 0.019 & 0.050 \\
(3c) & Bi-encoder$_{bert-base}$ & 500K \texttt{MSMarco-QA} & 0.038 & 0.098 & 0.043 & 0.113 & 0.014 & 0.040 \\
(3d) & Bi-encoder$_{mpnet-base}$ & 215M mul. sources & 0.138 & 0.297 & 0.108 & 0.277 & 0.023 & 0.076 \\
(3e) & Bi-encoder$_{mpnet-base}$ & 1.17B mul. sources & \textbf{0.155$^{\dagger 1}$} & \textbf{0.341$^{\dagger 12}$} & \textbf{0.147$^{\dagger 12}$} & \textbf{0.339$^{\dagger 12}$} & \textbf{0.041$^{\dagger}$} & \textbf{0.097$^{\dagger 12}$} \\ \midrule
 & \multicolumn{2}{l}{\textbf{Fine-tuned dense retrieval}} & \textbf{} & \textbf{} & \textbf{} & \textbf{} & \textbf{} & \textbf{} \\
 & \textbf{Model$_{LanguageModel}$} & \textbf{Negative sampler} &  &  &  &  &   \\ \midrule
(4a) & \multirow{3}{*}{Bi-encoder$_{mpnet-base}$ (3e)} & Random (0) & \textbf{0.130$^{\dagger}$} & \textbf{0.307$^{\dagger}$} & \textbf{0.168$^{\dagger 123}$} & \textbf{0.387$^{\dagger 123}$} & \textbf{0.050$^{\dagger 12}$} & \textbf{0.128$^{\dagger 123}$} \\
(4b) &  & \bm{} (1a) & 0.112 & 0.271 & 0.128 & 0.316 & 0.027 & 0.087 \\
(4c) &  & Bi-encoder (3e) & 0.065 & 0.146 & 0.144 & 0.306 & 0.018 & 0.051 \\ \bottomrule
\end{tabular}
\end{table*}

\section{Results}

In this section we first report on both dense and sparse retrieval results. Then we analyze the negative sampling procedure used to train the dense retrieval models.

\subsection{Sparse Retrieval}
In order to compare supervised and unsupervised sparse retrieval methods as well as zero-shot and fine-tuned dense retrieval models, we divided then into four categories as shown in Table~\ref{table:main_table_results}. Each row is a retrieval approach, containing the effectiveness in terms of R@1 and R@10 for each of the three datasets. 

\subsubsection*{\textbf{Does dialogue context expansion via RM3 lead to improvements over no expansion for sparse retrieval?}}
\bm{}+\rmthree{} (row 1b) does not improve over \bm{} (1a) on any of the three conversational datasets analyzed. A thorough hyperparameter fine-tuning was performed and no combination of the \rmthree{} hyperparameters outperformed \bm{}\footnote{Review appendix A for all hyperparameter combinations tested.}.  


A manual analysis of the new terms appended to a sample of 60 dialogue contexts reveals that only 18\% of them have at least one relevant term added based on our best judgement. Unlike web search where the query is often incomplete, under-specified, and ambiguous, in the information-seeking datasets employed in our work we observe that the dialogue context (query) is often times quite detailed and has more terms than the responses (documents). We hypothesize that because the dialogue contexts are already quite descriptive the task of expansion is trickier in this domain and thus we observe many dialogues for which the terms added are just noise.

\subsubsection*{\textbf{Does response expansion, i.e. \resptocontext{}, lead to improvements over no expansion for sparse retrieval?}}
We find that response expansion\footnote{See augmentation examples on Appendix B.} helped in two of the three datasets tested. \bm{}+\resptocontext{} (2a) outperforms \bm{} (1a) in two of the three datasets. 
Predicting only the last utterance of the dialogue (\resptocontextlu{}) performs better than predicting the whole utterance, as shown by \bm{}+\resptocontextlu{}'s (2b) higher recall values. For example in the \mantis{} dataset the R@10 goes from 0.309, when using the model trained to predict the dialogue context, to 0.325 when using the one trained to predict only the last utterance of the dialogue context.

In order to understand what the response expansion methods are doing most---term re-weighting or adding novel terms---we present the percentage of novel terms added by both methods in Table~\ref{table:resp2context_stats}. The table shows that \resptocontextlu{} does more term re-weighting than adding new words when compared to \resptocontext{} (53\% and 70\% on average are new words respectively and thus 47\% vs 30\% are changing the weights by adding existing words), generating overall smaller augmentations (115.45 vs 431.17 on average respectively).

In terms of sparse retrieval, the experiments so far reveal that using a response augmentation technique is a far better baseline than using BM25, which has been claimed to be a strong and proper baseline for comparison with dense models in conversational benchmarks~\cite{tao2021building,lan2021exploring}.

\begin{table}[ht!]
\caption{Statistics of the augmentations for the response (document) expansion methods \resptocontext{} and \resptocontextlu{}.}
\label{table:resp2context_stats}
\begin{tabular}{@{}lrrr@{}}
\toprule
 & \textbf{\mantis{}} & \textbf{\msdialog{} }& \textbf{\ubuntu{}} \\ \midrule
Context avg length & 474.12 & 426.08 & 76.95 \\
Response avg length & 42.58 & 71.38 & 11.06 \\ \hdashline
Aug. avg length - \resptocontext{} & 494.23 & 596.99 & 202.3 \\
Aug. avg length - \resptocontextlu{} & 138.5 & 135.29 & 72.57 \\
\% new words  -  \resptocontext{} & 71\% & 69\% & 71\% \\
\% new words  -  \resptocontextlu{} & 59\% & 37\% & 63\% \\ \bottomrule
\end{tabular}
\end{table}

\subsection{Dense Retrieval}

\subsubsection*{\textbf{Can zero-shot dense retrieval outperform a strong sparse baseline?}}

Zero-shot dense retrieval, i.e. no access to target data, beats the strong sparse baseline \bm{}+\resptocontext{} (2b) \emph{only when it is fine-tuned on large datasets containing diverse data including dialogues}, as we see by comparing rows (3a--c) and (3e--d) with row (2b) in Table~\ref{table:main_table_results}. For example, while the zero-shot dense retrieval models based only on the \texttt{MSMarco} dataset (3a--c) perform on average 35\% worse than the strong sparse baseline (2b) in terms of R@10 for the \msdialog{} dataset, the zero-shot model trained with 1.17B instances on diverse data (3e) is 68\% better than the strong sparse baseline (2b).

If the intermediate training data for the dense retrieval model is \texttt{MSMarco} (3a--c), none of the zero-shot retrieval models is able to beat the strong sparse baseline (2b).

When using however a bigger amount of intermediate training data, which includes Reddit and Stack Exchange responses\footnote{For the full description of the intermediate data see \url{https://huggingface.co/sentence-transformers/all-mpnet-base-v2}.}, we see that the zero-shot dense retrieval model (3e) is able to outperform the sparse retrieval baseline by margins of 33\% of R@10 on average across the datasets.

As expected, the closer the intermediate training data distribution is to the target domain, the better the dense retrieval model performs. The results indicate that a good zero-shot retrieval model needs to be trained for representation learning on a large set of datasets to outperform strong sparse retrieval baselines. Our results match previous empirical evidence on the effect of the intermediate training step on dense retrieval for different retrieval tasks~\cite{ouguz2021domain}.

\subsubsection*{\textbf{Is intermediate training of dense retrieval models helpful or is it sufficient to fine-tune a dense model on the target data?}}

Intermediate training on a large set of training instances is quite important for learning dense representations. Table~\ref{table:diff_lm_table} compares the dense models using either different pre-trained language models with and without using the intermediate data, with a different number of negative sampling procedures. 

We see that if we fine-tune \texttt{mpnet-base} directly on the target data, and do not do any intermediate training step the effectiveness drops are significant and substantial as shown when comparing results of 1.17B mul.sources (rows 1---3) vs no intermediate data (rows 4--6) in Table~\ref{table:diff_lm_table}. For example in the \mantis{} dataset the R@10 goes from 0.307 to 0.172 when using random negative sampling. This also happens for other language models and intermediate datasets, e.g. for \texttt{bert-base} and \texttt{MSMarco} the R@10 goes from 0.205 to 0.092 the \mantis{} dataset.

\subsubsection*{\textbf{What is the effect of fine-tuning the dense retrieval model after the intermediate training?}}

First, we see that simply using random sampling to find negatives and then fine-tune the dense retrieval model that had already gone through intermediate training--- row (4a) in Table~\ref{table:main_table_results}---achieves the best overall effectiveness we obtain in two of the three datasets. Having access to the target conversational data as opposed to only a diverse set of question and answers means that the representations learned by the model are closer to the true distribution of the data.

We hypothesize that fine-tuning Bi-encoder$_{mpnet-base}$ (3e) for \mantis{} (4a) is harmful because the intermediate data contains multiple Stack Exchange responses. In this way, the subset of dialogues of Stack Exchange that \mantis{} encompasses might be serving only to overfit the intermediate representations. As evidence for this hypothesis, we found that (I) the learning curves flatten quickly (as opposed to other datasets) and (II) fine-tuning another language model that does not have Stack Exchange data (\texttt{MSMarco}) in their fine-tuning, Bi-encoder$_{bert-base}$ (3c), improves the effectiveness with statistical significance from 0.092 R@10 to 0.205 R@10, as shown in Table~\ref{table:diff_lm_table}.

\subsubsection*{\textbf{Do harder negative samples lead to more effective fine-tuning of dense retrieval models?}}

Surprisingly we found that using more effective models to select negative candidates is detrimental to the effectiveness of the dense retrieval model (rows 4a--c). We observe this phenomena when using different different language models and whether using intermediate training or not for all datasets tested, as shown in Table~\ref{table:diff_lm_table}.

\begin{figure}[]
    \centering
    \includegraphics[width=.48\textwidth]{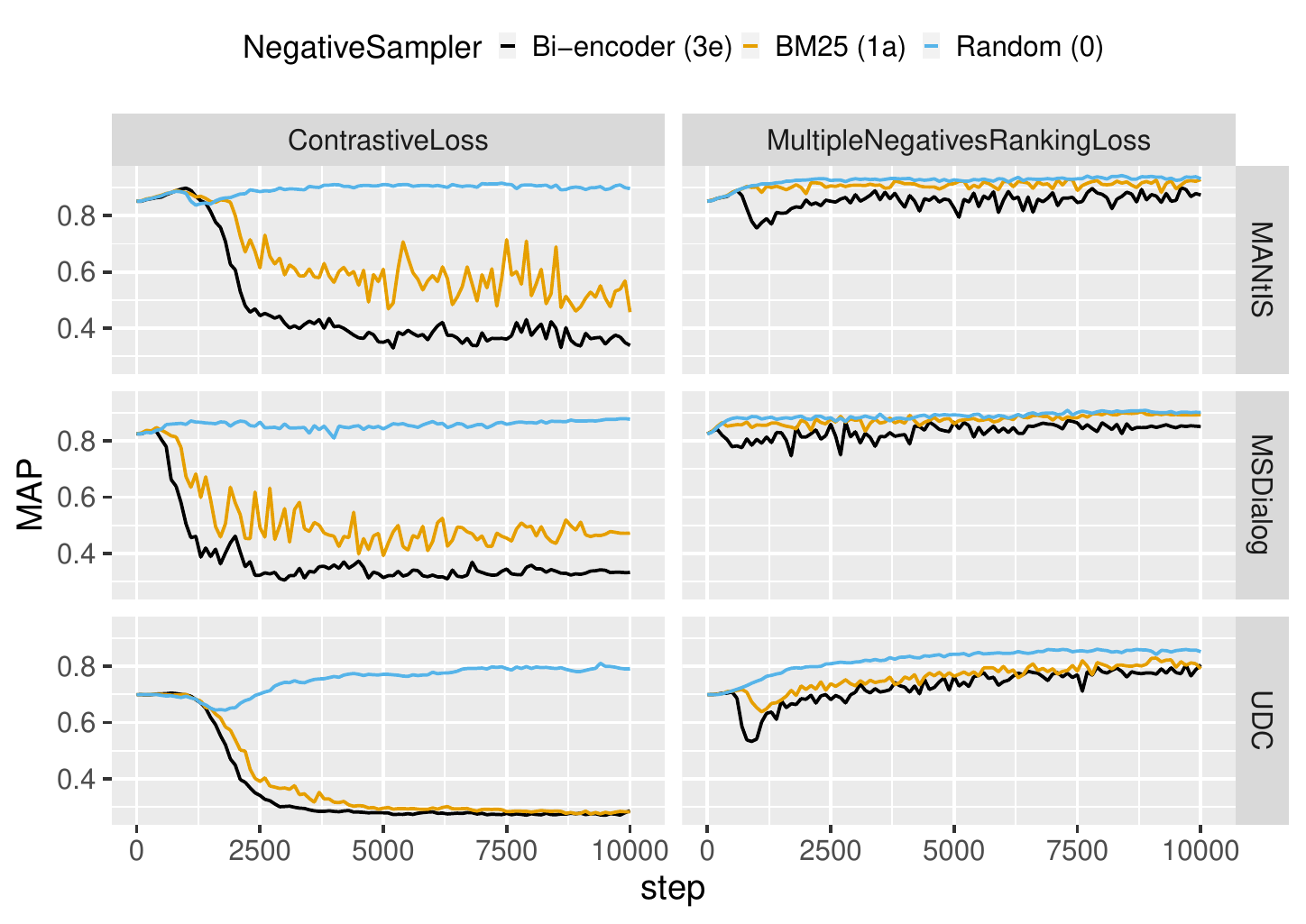}
    \caption{Validation effectiveness training curves for the re-ranking task (10 candidates obtained using random sampling). The plot displays the dense retrieval Bi-encoder$_{mpnet-base}$ model when using different loss functions for each negative sampling procedures (the right column is the training curves for rows 4a--c on Table~\ref{table:main_table_results}). We observe the phenomena of decreasing effectiveness for better negative sampling methods when using different loss functions.}
    \label{fig:diff_loss}
\end{figure}

Figure~\ref{fig:diff_loss} shows validation curves during training for the \textit{MultipleNegativesRankingLoss} and an alternative contrastive loss~\cite{hadsell2006dimensionality} (\textit{ContrastiveLoss}) that does employ in-batch negative sampling. The experiment tries to isolate the effect the specific loss function might be having on the negative sampling phenomena, and we observe that the same behaviour regardless of the loss function\footnote{Other loss functions besides the two plotted were also tested and resulted in the the same effectiveness for the negative samplers: $Random >> \bm{} >> Bi-encoder$.}. 

Based on brainstorm sessions and discussions the authors had with other IR researchers a set of hypotheses were formed that could explain why this phenomena might be happening. Next we explore the three resulting hypotheses with six additional experiments.

\begin{table*}[]
\caption{Effectiveness of fine-tuned dense retrieval models when using different language models and intermediate training for each negative sampling procedures from Table~\ref{table:main_table_results}. Bold indicates the highest value within different negative sampling methods for the same setting. We observe the same phenomena of decreasing effectiveness for better negative sampling methods when using different language models and whether using intermediate training or not.}
\label{table:diff_lm_table}
\begin{tabular}{@{}llrrrrrrr@{}}
\toprule
 &  &  & \multicolumn{2}{c}{\textbf{\mantis{}}} & \multicolumn{2}{c}{\textbf{\msdialog{}}} & \multicolumn{2}{c}{\textbf{\ubuntu{}}} \\ \midrule
\textbf{Model} & \textbf{Intermediate data} & \textbf{Neg. Sampler} & \textbf{R@1} & \textbf{R@10} & \textbf{R@1} & \textbf{R@10} & \textbf{R@1} & \textbf{R@10} \\ \midrule
\multirow{6}{*}{Bi-encoder$_{mpnet-base}$} & \multirow{3}{*}{1.17B mul.sources} & Random (0) & \textbf{0.130} & \textbf{0.307} & \textbf{0.168} & \textbf{0.387} & \textbf{0.050} & \textbf{0.128} \\
 &  & \bm{} (1a) & 0.112 & 0.271 & 0.128 & 0.316 & 0.027 & 0.087 \\
 &  & Bi-encoder (3e) & 0.065 & 0.146 & 0.144 & 0.306 & 0.018 & 0.051 \\ \cmidrule(l){2-9} 
 & \multicolumn{1}{c}{\multirow{3}{*}{-}} & Random (0) & \textbf{0.070} & \textbf{0.172} & \textbf{0.114} & \textbf{0.308} & \textbf{0.021} & \textbf{0.063} \\
 & \multicolumn{1}{c}{} & \bm{} (1a) & 0.043 & 0.118 & 0.091 & 0.256 & 0.009 & 0.027 \\
 & \multicolumn{1}{c}{} & Bi-encoder (3e) & 0.032 & 0.087 & 0.083 & 0.205 & 0.002 & 0.019 \\ \midrule
\multirow{6}{*}{Bi-encoder$_{bert-base}$} & \multirow{3}{*}{500K \texttt{MSMarco-QA}} & Random (0) & \textbf{0.085} & \textbf{0.205} & \textbf{0.138} & \textbf{0.339} & \textbf{0.030} & \textbf{0.079} \\
 &  & \bm{} (1a) & 0.051 & 0.130 & 0.116 & 0.287 & 0.007 & 0.022 \\
 &  & Bi-encoder (3e) & 0.043 & 0.106 & 0.107 & 0.242 & 0.008 & 0.030 \\ \cmidrule(l){2-9} 
 & \multicolumn{1}{c}{\multirow{3}{*}{-}} & Random (0) & \textbf{0.029} & \textbf{0.092} & \textbf{0.063} & \textbf{0.200} & \textbf{0.012} & \textbf{0.038} \\
 & \multicolumn{1}{c}{} & \bm{} (1a) & 0.017 & 0.057 & 0.040 & 0.144 & 0.002 & 0.006 \\
 & \multicolumn{1}{c}{} & Bi-encoder (3e) & 0.011 & 0.041 & 0.034 & 0.119 & 0.000 & 0.009 \\ \bottomrule
\end{tabular}
\end{table*}

\subsection{Dense Retrieval: Negative Sampling}

We investigated the following hypotheses that could explain the observed phenomena of decreasing effectiveness for better negative sampling functions:

\begin{itemize}
    \item[\textbf{H1}] False negative samples increase when using better negative sampling methods. False negative are responses that are potentially valid for the context. Such relevant responses sampled will lead to unlearning relevant matches between context and responses as they receive negative labels. Example retrieved by the Bi-encoder model (line 3e of Table~\ref{table:main_table_results}):
    
    \vspace{3mm} 
     \fbox{
     \begin{minipage}{23em}
    \begin{itemize}[leftmargin=-0.1cm] 
        \item[] \textbf{Dialogue context ($\set{U}$)}: {\color{BlueViolet} hey... how long until dapper comes out?} $\usep$ {\color{purple}14 days} [...] $\usep$ {\color{BlueViolet}i thought it was coming out tonight}
        \item[] \textbf{Correct response ($r^{+}$)}: {\color{purple}just kidding couple hours}
        \item[] \textbf{Negative sample ($r^{-}$)}: there is a possibility dapper will be delayed [...] meanwhile, dapper discussions should occur in ubuntu+1
    \end{itemize}
    \end{minipage}}
    \vspace{3mm} 
    
    \item[\textbf{H2}] Confusing negative samples increase when using better negative sampling methods. They are not relevant, i.e. a valid response to the context, but they are semantically or lexically identical to (or exact matches or part of) the context. Such negative samples will lead to representations of similar sentences to be far apart in the embedding space. Example of a partial match retrieved by BM25 (line 1a of Table~\ref{table:main_table_results}):
    
    \vspace{3mm} 
     \fbox{\begin{minipage}{23em}
    \begin{itemize}[leftmargin=-0.1cm] 
        \item[] \textbf{Dialogue context ($\set{U}$)}: {\color{BlueViolet}can any one help me im trying to install some thing and i get this error GTK... configure: error: Package requirements (gtk+-2.0} $\usep$ {\color{purple}perhaps \underline{sudo apt-get install libgtk2.0-dev}} $\usep$ {\color{BlueViolet}any way to tell it to install all dependencies too}
        \item[] \textbf{Correct response ($r^{+}$)}: {\color{purple}what do you mean, it won't compile if you don't have the dependencies}
        \item[] \textbf{Negative sample ($r^{-}$)}: sudo apt-get install libgtk2.0-dev
    \end{itemize}
    \end{minipage}}
    \vspace{3mm}

    \item[\textbf{H3}] There is a lack of informative negative samples, i.e. responses that are more informative than random negative responses for training, for the dialogue contexts within each dataset. Informative negative samples are ideally the ones that (I) have lexical matches with the dialogue context but are not subsets of continuous parts of the context and are not semantically relevant or (II) are potential responses to the dialogue (gives the impression that it is a natural and fluent response to the last utterance of the dialogue context) but are not semantically relevant.
    
    Unlike a web collection where we can likely find the same term used in different contexts, the dialogue collections used in this work have only a few domains, and a single term is less likely then to appear within different contexts. Moreover, different to common adhoc retrieval tasks such as web search and passage ranking, in response ranking the response (negative sample) has to be a likely continuation of the last utterance of the dialogue context. Examples of potentially informative negative samples:
    
    \vspace{3mm} 
     \fbox{\begin{minipage}{23em}
    \begin{itemize}[leftmargin=-0.1cm] 
        \item[] \textbf{Dialogue context ($\set{U}$)}: {\color{BlueViolet}I had my iPhone swapped out by Apple and after reinstalling my apps, signing in, etc, I noticed my OneDrive app was saying ""Be sure you're connected to cellular or wifi"".... and it is. I've signed out and back in... removed and re added the app... etc no dice. Anyone have any suggestions?}

        \item[] \textbf{Correct response ($r^{+}$)}: {\color{purple}Hi PERSON\_PLACEHOLDER, I realized the inconvenience you are experiencing. I certainly help you. Is the issue specific to OneDrive app or with other apps as well? First, update iOS on your device. Then, make sure you've installed any available updates to the app. [...]}

        \item[] \textbf{Different collection negative sample ($r^{-}$)\footnote{Response selected from \textit{reddit/r/onedrive} dialogues.}}: I love OneDrive, have used it for years with no issues. I believe a lot of people have issues because they don't understand how it works, they don't read the instructions and when they install it they just click 'yes' to everything.

        \item[] \textbf{Generated negative sample ($r^{-}$)}\footnote{Response generated by \textit{DialoGPT-large} for the dialogue context.}: I had the same problem. I had to uninstall and reinstall the app.

    \end{itemize}
        \end{minipage}}
    \vspace{3mm}
    
\end{itemize}

In order to test our hypotheses we perform the following experiments, each one geared towards investigating one hypothesis:

\begin{itemize}
    \item[\textbf{E1}] Manually annotate a subset of negative samples in terms of their relevance, to check whether the number of false negatives increases with better negative sampling functions (\textbf{H1}).
    \item[\textbf{E2}] Instead of using the top ranked responses as negative responses, we use the bottom responses of the top ranked responses as negatives. As an example, when we retrieve $k=100$ responses, instead of using responses ranked 1 to 10 we use responses ranked 91 to 100. This decreases the chances of obtaining false positives and if we set $k$ to small values such as 100, it will not render the sampling procedure to be essentially random (\textbf{H1}).
    \item[\textbf{E3}] Remove negative samples that are subsets of the context when training dense models and compare its effectiveness with the original negative samples (\textbf{H2}).
    \item[\textbf{E4}] Use only the last utterance to retrieve negative samples, this will make it less likely that a response is an exact match with the entire dialogue context (\textbf{H2}).
    \item[\textbf{E5}] Compare the effectiveness of dense retrieval models when using a corpus of responses for negative sampling which has additional responses from external corpora, that are potentially more informative than the ones from the original dataset (\textbf{H3}).
    \item[\textbf{E6}] Generate negative samples for the contexts using a generative language model and compare the effectiveness of this model against using retrieved negative samples (\textbf{H3}).
\end{itemize}

\begin{table}[]
\small
\caption{Experiments to examine why better negatives sampling procedures lead to worse dense retrieval results. Bold indicates positive evidence for the corresponding hypothesis. We present the R@10 gains between the condition presented and the absence of the condition (4a--c on Table~\ref{table:main_table_results}) on E2--E5.}
\label{table:hypothesis_res}
\begin{tabular}{@{}llccc@{}}
\toprule
\multicolumn{5}{c}{E1} \\ \midrule
NS & Count & \mantis{} & \msdialog{} & \ubuntu{} \\ \midrule
Random (0) & \multirow{2}{*}{false $r^{-}$} & 0 & 0 & 0 \\
\bm{} (1a) &  & 0 & \textbf{4} & \textbf{2} \\
Bi-encoder (3e) &  & \textbf{11} & \textbf{4} & \textbf{15} \\ \midrule
\multicolumn{5}{c}{E2} \\ \midrule
 &  & \mantis{} & \msdialog{} & \ubuntu{} \\ \midrule
NS & Condition & R@10$\Delta$ & R@10$\Delta$ & R@10$\Delta$ \\ \midrule
\bm{} (1a) & \multirow{2}{*}{denoising} & -0.014 & \textbf{+0.042} & \textbf{+0.034} \\
Bi-encoder (3e) &  & \textbf{+0.170} & \textbf{+0.091} & \textbf{+0.056} \\ \midrule
\multicolumn{5}{c}{E3} \\ \midrule
 &  & \mantis{} & \msdialog{} & \ubuntu{} \\ \midrule
NS & Condition & R@10$\Delta$ & R@10$\Delta$ & R@10$\Delta$ \\ \midrule
\bm{} (1a) & \multirow{2}{*}{$r^{-}$ subset of $\set{U}$} & \textbf{-0.032} & +0.015 & \textbf{-0.010} \\
Bi-encoder (3e) &  & +0.034 & \textbf{-0.040} & \textbf{-0.003} \\ \midrule
\multicolumn{5}{c}{E4} \\ \midrule
 &  & \mantis{} & \msdialog{} & \ubuntu{} \\ \midrule
NS & Condition & R@10$\Delta$ & R@10$\Delta$ & R@10$\Delta$ \\ \midrule
\bm{} (1a) & \multirow{2}{*}{U$_{lu}$ as query} & -0.002 & \textbf{+0.044} & -0.008 \\
Bi-encoder (3e) &  & \textbf{+0.173} & \textbf{+0.042} & \textbf{+0.047} \\ \midrule
\multicolumn{5}{c}{E5} \\ \midrule
 &  & \mantis{} & \msdialog{} & \ubuntu{} \\ \midrule
NS & Corpus & R@10$\Delta$ & R@10$\Delta$ & R@10$\Delta$ \\ \midrule
\bm{} (1a) & \multirow{2}{*}{expanded} & -0.014 & \textbf{+0.031} & \textbf{+0.023} \\
Bi-encoder (3e) & & \textbf{+0.113} & \textbf{+0.028} & \textbf{+0.057}   \\ \midrule
\multicolumn{5}{c}{E6} \\ \midrule
 &  & \mantis{} & \msdialog{} & \ubuntu{} \\ \midrule
\multicolumn{2}{l}{NS} & R@10 & R@10 & R@10 \\ \midrule
\multicolumn{2}{l}{Random (0)} & 0.307 & 0.387 & 0.128 \\
\multicolumn{2}{l}{GenNegatives$_{Blenderbot}$} & 0.267 & 0.348 & \textbf{0.134} \\
\multicolumn{2}{l}{GenNegatives$_{DialoGPT}$} & 0.260 & 0.363 & 0.123 \\ \bottomrule
\end{tabular}
\end{table}

\begin{table}[]
\caption{Summary of the findings of our follow-up experiments to explain why better negative samplers lead to worse dense retrieval results.}
\label{table:evidence_for_h}
\begin{tabular}{@{}ccc@{}}
\toprule
\multicolumn{1}{l}{Experiment} & \multicolumn{1}{l}{Hypothesis} & \multicolumn{1}{l}{Evidence for H} \\ \midrule
E1 & \multirow{2}{*}{H1} & \cmark \\
E2 &  & \cmark \\ \midrule
E3 & \multirow{2}{*}{H2} & \xmark \\
E4 &  & \cmark \\ \midrule
E5 & \multirow{2}{*}{H3} &  \cmark \\
E6 &  & \xmark \\ \bottomrule
\end{tabular}
\end{table}

Our findings for the six experiments (E1--E6) are displayed in Table~\ref{table:hypothesis_res}. Bold values indicate positive evidence for their respective hypothesis. 

\textbf{In the first experiment (E1)}, we manually annotated the relevance of 270 pairs of dialogue context and negative samples (3 datasets $\times$ 3 dialogue contexts $\times$ 10 negative samples $\times$ 3 negative sampling method). We found that indeed the number of false positives increases when using better negative sampling approaches, providing positive evidence for the hypothesis that false positives are detrimental to the training of the dense retrieval models. \textbf{For the second experiment (E2)} we employ a denoising technique that uses the bottom negative samples from the top-k list instead of the first. We found that the effectiveness improves by large margins when using dense model to find negatives in all three datasets. In two datasets (\mantis{} and \msdialog{}) we find that the denoised negative sampling of the Bi-encoder yields statistically significant improvements over Random (0.316 R@10 vs 0.307 R@10 for \mantis{} and 0.397 R@10 vs 0.387 for \msdialog{}). The results for the second experiment are thus additional positive evidence for the hypothesis that false positives are detrimental. 

\textbf{In the third experiment (E3)}, by allowing the negative samples to be subsets of the dialogue context, we expect the effectiveness of the model to drop by large margins since the number of confusing negative samples increase. This was not the case. The results indicates that possibly confusing negative samples with exact matches with the dialogue context were not detrimental. \textbf{For the fourth experiment (E4)} we expected that when using only the last utterance of the dialogue to find negatives, we would decrease the number of confusing negatives. This was the case for training the model with the bi-encoder as the negative sampler.

In the final two experiments we tested whether we could find more informative samples \textbf{by using an expanded corpus of responses (E5) and by using generated negative responses (E6)}. We found that using the larger corpus was beneficial when using the bi-encoder negative sampler, showing that we can possibly find more informative negative samples when using larger data. We found however that the generated negative responses from both models were not effective, as random samples from lead to better effectiveness when training the dense retrieval model.

A summary interpretation of the experiments can be found in Table~\ref{table:evidence_for_h}, where for each experiment we display whether we found evidence or not for its hypothesis. Overall we see that we have the most evidence for the first hypothesis (H1) of false negatives degrading the training procedure. The problems of false negatives when using harder negatives has been discussed before for other retrieval tasks~\cite{qu2020rocketqa,gao2021unsupervised}, and we find evidence here on the conversational task that matches prior works on denoising the hard negatives. Other hypothesis (H2 and H3) had partial positive evidence, which suggests that they could also be a potential source of difficulty when training dense models with harder negatives. 

In conclusion, we demonstrate that a denoising strategy to remove false negative samples is required to train dense models for ranking responses for conversations, when taking into account hard negative samples.    
\section{Conclusion}
We study here the problem of full-rank retrieval of responses for dialogues. We explore sparse and dense techniques that retrieve responses out of the entire collection available. The expansion of responses, i.e. \resptocontextlu{} showed to be a strong baseline for sparse retrieval. We also find that dense retrieval needs large datasets in order to beat a strong sparse retrieval baseline in the zero-shot setting. Our findings also suggest that fine-tuning a bi-encoder dense retrieval model after intermediate training to be the best performing method for the task of full-rank retrieval of responses for dialogues. We finish our experiments with a thorough analysis of negative sampling methods, exploring hypothesis that could explain why harder negatives lead to worse effectiveness for the dense methods.

As future work we believe important directions include: taking advantage of language-model based term re-weighting for sparse retrieval, intermediate training strategies in order to improve their generalization power as well as dense-sparse hybridization techniques for the full-rank retrieval of responses for dialogues.

\begin{acks}
This research has been supported by NWO projects SearchX (639.022.722) and NWO Aspasia (015.013.027).
\end{acks}

\bibliographystyle{ACM-Reference-Format}
\bibliography{references}

\appendix

\section{RM3 hyperparameters tested}
\begin{table*}[hbt!]
\small
\caption{\textcolor{Brown}{Appendix A.} Hyperparameters tested for RM3. The hyperparameters are: number of expansion terms - number of expansion documents - weight to assign to the original query. We see that when we use more of the original query we get higher effectiveness and other hyperparameters do not help much either decreasing or increasing.}
\label{table:rm3_results}
\begin{tabular}{@{}lllllll@{}} %
\toprule
                               & \multicolumn{2}{l}{\textbf{\mantis{}}} & \multicolumn{2}{l}{\textbf{\msdialog{}}} & \multicolumn{2}{l}{\textbf{\ubuntu{}}} \\ \midrule
                               & \textbf{R@1}       & \textbf{R@10}     & \textbf{R@1}        & \textbf{R@10}      & \textbf{R@1}       & \textbf{R@10}     \\ \midrule
\bm{} & \textbf{0.133} & \textbf{0.299} & \textbf{0.064} & \textbf{0.177} & \textbf{0.027} & \textbf{0.070} \\ \hdashline
+\rmthree{} (5-5-0.5) & 0.089 & 0.214 & 0.042 & 0.133 & 0.014 & 0.042 \\
+\rmthree{} (5-5-0.7) & 0.097 & 0.238 & 0.048 & 0.150 & 0.017 & 0.051 \\
+\rmthree{} (5-10-0.5) & 0.072 & 0.190 & 0.036 & 0.123 & 0.010 & 0.037 \\
+\rmthree{} (5-10-0.7) & 0.086 & 0.218 & 0.043 & 0.145 & 0.012 & 0.047 \\
+\rmthree{} (5-15-0.5) & 0.066 & 0.175 & 0.032 & 0.115 & 0.008 & 0.038 \\
+\rmthree{} (5-15-0.7) & 0.080 & 0.206 & 0.039 & 0.135 & 0.010 & 0.047 \\
+\rmthree{} (10-5-0.5) & 0.094 & 0.248 & 0.042 & 0.150 & 0.016 & 0.053 \\
+\rmthree{} (10-5-0.7) & 0.103 & 0.266 & 0.048 & 0.161 & 0.019 & 0.058 \\
+\rmthree{} (10--10-0.5) & 0.081 & 0.221 & 0.038 & 0.136 & 0.014 & 0.049 \\
+\rmthree{} (10-10-0.7) & 0.092 & 0.246 & 0.046 & 0.153 & 0.015 & 0.057 \\
+\rmthree{} (10-15-0.5) & 0.073 & 0.206 & 0.035 & 0.127 & 0.011 & 0.049 \\
+\rmthree{} (10-15-0.7) & 0.085 & 0.232 & 0.044 & 0.142 & 0.017 & 0.056 \\
+\rmthree{} (15-5-0.5) & 0.099 & 0.261 & 0.045 & 0.152 & 0.019 & 0.057 \\
+\rmthree{} (15-5-0.7) & 0.107 & 0.275 & 0.050 & 0.162 & 0.022 & 0.061 \\
+\rmthree{} (15--10-0.5) & 0.085 & 0.238 & 0.036 & 0.145 & 0.013 & 0.054 \\
+\rmthree{} (15-10-0.7) & 0.097 & 0.258 & 0.048 & 0.156 & 0.016 & 0.059 \\
+\rmthree{} (15-15-0.5) & 0.077 & 0.219 & 0.036 & 0.131 & 0.014 & 0.054 \\
+\rmthree{} (15-15-0.7) & 0.091 & 0.245 & 0.042 & 0.148 & 0.017 & 0.060 \\ \bottomrule
\end{tabular}
\end{table*}

\section{Examples of \resptocontext{} augmentations}

\begin{table*}
\small
\caption{\textcolor{Brown}{Appendix B.} Random examples of augmentations by the \resptocontext{} methods.}
\label{table:resp2contex_examples}
\begin{tabular}{@{}p{1cm}p{6.5cm}p{2.5cm}p{3cm}p{3cm}@{}}
\toprule
Dataset & Context & Response & Augmentation \resptocontext{} & Augmentation \resptocontextlu{} \\ \midrule
\mantis{} & On macOs High Sierra I am trying to set a default icon for .pdf files: when I used Adobe Reader all the pdf files showed its icon, then I uninstalled it and replaced it with Foxit Reader as default app to open the same file. However all the pdf files now show a white icon and I would like them to show the Foxit logo instead.I tried to copy-paste the logo from Foxit to a random pdf and it does work, but this cannot be applied to the whole configuration (this means that I should copy-paste everytime the logo in every single pdf file).Is there a way to set the default icon? [UTT\_SEP] "Once you have changed the default application using Get Info, you may need to clear the icon cache.To get started, close all open applications and then launch Terminal from Macintosh HD > Applications > Utilities. Copy and paste the following command at the Terminal prompt and press Return to execute it: [code removed for paper] Restart, job done.From [URL] Safari download bar rar file icon show as VLC icon,which associate to The Unarchiver should shows its icon" [UTT\_SEP] I did what you explained but the problem still exist. I have updated to Mojave since the last time. In Finder (choosing the list view for my files) I still see .pdf icon represented as a blank page. & It appears Foxit Reader is missing the readerpdf.icns file. As a quick fix until the developer releases a solution you could borrow the PDF icon from another app or find an icon from a previous version. & The newest version of Foxit Reader is , which comes with Firefox, and has an icon for Adobe PDF Reader.  is a version that was released by Foxit in May of 2014. However, this is not an update, it has been updated to a different version.  is in a separate file that the reader PDF icon can be found in a new window and it is a  PDF icon a new icon is found on the new window and not the app store. I have not been [...] & I tried it and it worked. I just reinstalled the software and the app works fine. I have to say that i can't find a PDF icon. I've renamed and downloaded the app. \\ \midrule
\msdialog{} & KF-UK you are not alone. What I did, is I uninstalled all new updates and paused the download feature. Hopefully someone has a workaround before serious problems occur. As you know Microsoft is a security conscious organization looking out for the well being of it customers. In other words we need your money but if the matter hits the fan we have no knowledge of your existence but would you recommend our products to your friends.   [UTT\_SEP]  Even after KB4040724 (build 15063.632) update, the issue persists (in my case). There are two workaround for this issue reguarding my opinion at the moment: 1 - at login or later, the combination of shift-ctrl-win + B restores the graphics card and this ""awaken"" the monitor in standby 2 - Turn off Fast Boot by disabling it via Control Panel \ Power Options \ Choose What the Power Buttons do \ Change Settings that are Currently Unavailable \ untick ""Turn on fast startup"". & I too am having problems with a dual monitor setup. My system had KB4038788 installed this month. I will attempt an uninstall and see if the dual monitor sleep issue goes away.... Complete symptom details for me are below. I am subscribing to this thread hoping Microsoft provides an answer. & My 870 has been working great and my 810 is having some problems (like the aforementioned one). Is this an update that is supposed to fix the 870, which I'm not sure I'm missing - I'm not sure what happened and if I'm not a 870, I'm not a 870. I don't know what happened and it just hangs up in bed [...] & for your reply. I have the same problem as you, if I am on my iMac. When I wake up, it will go to sleep but when I get up to do that the same thing happens again. It's also not the same. I'm using the 2nd monitor. I've also slept well on this monitor for 2 years, so it' \\ \midrule
\ubuntu{} & anyone notices there is something wrong with man page of netstat..... [UTT\_SEP] wondering if its only me... or there is something wrong [UTT\_SEP] it looks fine to me [UTT\_SEP] my man page is not showing argument options, like -an etc & mine shows both -a and -n :\ & Hello, a lot of ubuntu users have asked me this on the forum, and I just can't seem to figure this out, but when I open my terminal, it asks me ""Can I do anything with my ubuntu"?" [UTT\_SEP] how [...] & how to change the name of a pc in gnome-server? \\ \bottomrule
\end{tabular}
\end{table*}

\end{document}